\pdfoutput=1
\documentclass[sigconf]{acmart}
\usepackage[utf8]{inputenc}

\usepackage{algorithmic}
\usepackage{cleveref}
\usepackage{comment}
\usepackage{lipsum}
\usepackage{listings}
\usepackage{graphicx}
\usepackage{siunitx}
\usepackage{subfigure}
\usepackage{textcomp}
\usepackage{xcolor}
\usepackage{xspace}

\graphicspath{{images/}{pdfs/}{figs/}{svgs/}{odgs/}{figures/}}
\DeclareGraphicsExtensions{.fig,.pdf,.png,.jpg}

\theoremstyle{definition}
\newtheorem{researchquestion}{RQ}
\crefname{researchquestion}{RQ}{RQs}

\newcommand{\SWH}{Software Heritage\xspace}

\usepackage[textsize=tiny]{todonotes}

 \copyrightyear{2026}
\acmYear{2026}
\setcopyright{cc}
\setcctype{by}
\acmConference[ACM REP '26]{ACM Conference on Reproducibility and Replicability}{July 20--22, 2026}{Delft, Netherlands}
\acmBooktitle{ACM Conference on Reproducibility and Replicability (ACM REP '26), July 20--22, 2026, Delft, Netherlands}
\acmDOI{10.1145/3820002.3828585}
\acmISBN{979-8-4007-2778-8/2026/07}

\newcommand{\DatasetDate}{October 8, 2025\xspace}
\newcommand{\CutoffDate}{February 23, 2016\xspace}
\newcommand{\TemporalSpan}{2016--2025\xspace}

\newcommand{\TotalOriginsAnalyzedShort}{\num{328.4}\,M\xspace}

\newcommand{\OriginsFilteredOutShort}{\num{74.4}\,M\xspace}

\newcommand{\OriginsWithMultipleSnapshotsShort}{\num{328.4}\,M\xspace}

\newcommand{\OriginsWithTagsShort}{\num{45.5}\,M\xspace}

\newcommand{\TotalAlterations}{\num{10 171 046}\xspace}
\newcommand{\TotalAlterationsShort}{\num{10.2}\,M\xspace}
\newcommand{\AlteredRepositories}{\num{188 713}\xspace}
\newcommand{\AlteredRepositoriesShort}{\num{189}\,k\xspace}
\newcommand{\AlteredRepositoriesPercent}{\num{0.41}\%\xspace}
\newcommand{\AlteredTagPerRepository}{\num{53.9}\xspace}
\newcommand{\AlteredTagPerRepositoryMedian}{\num{3}\xspace}

\newcommand{\NonLegitAlterationsShort}{\num{5}\,M\xspace}

\newcommand{\LegitMoves}{\num{626 008}\xspace}

\newcommand{\Deletions}{\num{9 545 038}\xspace}

\newcommand{\DeletionsPercent}{\num{93.85}\%\xspace}
\newcommand{\MovesWithContentChange}{\num{426 595}\xspace}

\newcommand{\PercentMovesContentChange}{\num{68.15}\%\xspace}

\newcommand{\ReleaseDiff}{\num{10 439}\xspace}

\newcommand{\ReleaseDiffPercent}{\num{1.67}\%\xspace}

\newcommand{\RevisionDiff}{\num{188 974}\xspace}

\newcommand{\RevisionDiffPercent}{\num{30.19}\%\xspace}

\newcommand{\StarRangeZeroOne}{0--1\xspace}
\newcommand{\StarRangeTwoTen}{2--10\xspace}
\newcommand{\StarRangeElevenHundred}{11--100\xspace}
\newcommand{\StarRangeHundredFiveHundred}{101--500\xspace}
\newcommand{\StarRangeFiveHundredPlus}{501+\xspace}

\newcommand{\StarZeroOneAlterations}{\num{7 555 751}\xspace}
\newcommand{\StarZeroOneAlterationsPercent}{\num{74.29}\%\xspace}
\newcommand{\StarZeroOneOrigins}{\num{103 453}\xspace}
\newcommand{\StarZeroOneOriginsPercent}{\num{54.82}\%\xspace}

\newcommand{\NixPkgsTracked}{\num{2 407}\xspace}
\newcommand{\NixPkgsAffected}{\num{32}\xspace}
\newcommand{\NixPkgsHashMismatch}{\num{7}\xspace}

\newcommand{\MinSnapshotsRequired}{2\xspace}

\title[Mutating the ``Immutable'': A Large-Scale Study of Git Tag Alterations]{Mutating the ``Immutable'':\\A Large-Scale Study of Git Tag Alterations}
\thanks{This work was supported by France Agence Nationale de la Recherche (ANR), program France 2030, reference ANR-22-PTCC-0001. This work was made possible by Software Heritage, the universal source code archive:\linebreak \url{https://www.softwareheritage.org}.}
\author{Solal Rapaport}
\email{solal.rapaport@telecom-paris.fr}
\orcid{0009-0007-2083-0254}
\affiliation{\institution{LTCI, Télécom Paris, Institut Polytechnique de Paris}
  \city{Palaiseau}
  \country{France}
}

\author{Laurent Pautet}
\email{laurent.pautet@telecom-paris.fr}
\affiliation{\institution{LTCI, Télécom Paris, Institut Polytechnique de Paris}
  \city{Palaiseau}
  \country{France}
}

\author{Samuel Tardieu}
\email{samuel.tardieu@telecom-paris.fr}
\affiliation{\institution{LTCI, Télécom Paris, Institut Polytechnique de Paris}
  \city{Palaiseau}
  \country{France}
}

\author{Stefano Zacchiroli}
\email{stefano.zacchiroli@telecom-paris.fr}
\orcid{0000-0002-4576-136X}
\affiliation{\institution{LTCI, Télécom Paris, Institut Polytechnique de Paris}
  \city{Palaiseau}
  \country{France}
}

\author{Théo Zimmermann}
\email{theo.zimmermann@telecom-paris.fr}
\affiliation{\institution{LTCI, Télécom Paris, Institut Polytechnique de Paris}
  \city{Palaiseau}
  \country{France}
}

\begin{abstract}

Git tags are commonly viewed as immutable references in software development, marking releases and specific repository states that underpin build reproducibility and software supply-chain integrity.
Despite their intended immutability, Git allows tags to be altered through deletion or modification via force-pushed updates.
The prevalence of such alterations threatens reproducible builds and dependency integrity.

We conduct the first large-scale empirical study of tag alterations in public code repositories, analyzing \TotalOriginsAnalyzedShort software repositories from \SWH and identifying \TotalAlterationsShort tag alterations affecting \AlteredRepositoriesShort unique repositories.
A cross-analysis with Nixpkgs reveals that \NixPkgsAffected packages reference tags altered in our dataset, with \NixPkgsHashMismatch exhibiting confirmed build errors, providing concrete evidence that tag alterations break reproducible package builds.

Our findings challenge the widespread assumption that tags are immutable anchors for released software.
We therefore recommend that build systems and package managers pin dependencies to cryptographic commit hashes, that development forges expose tag-mutation audit logs, and that the community adopt systematic monitoring of tag alterations as a standard supply-chain security practice.

\end{abstract}


\begin{document}
\maketitle

\section{Introduction}
\label{sec:intro}

Git tags are fundamental constructs in modern software development, serving as immutable references for specific points in repository history~\cite[§2.6, Tagging]{ProGit2014}.
They play a significant role in establishing stable references for software releases, enabling reproducible software environments and builds, and anchoring dependency management systems.
Dependency management ecosystems can rely on Git tags to resolve and pin specific versions.
Most notably, Go modules, as well as package managers like npm (via \texttt{git://} URLs in \texttt{package.json}), Cargo (via \texttt{git} dependencies in \texttt{Cargo.toml}), and pip, support direct Git repository dependencies anchored to tags.
Continuous integration systems (\emph{e.g.}, GitHub Actions) also use them to anchor deployment pipelines.

However, Git's design permits tag alterations---
both lightweight (implemented as ``refs'' in Git terminology) and annotated tags can be deleted or force-updated to point to different commits.
While such modifications may serve legitimate purposes (correcting release errors, repository cleanup, or repository migrations), they violate the immutability assumption that downstream tooling and practices depend upon.

The mutability of Git tags has profound implications for both software security and reproducibility.
Attackers can exploit tag mutability to inject malicious code into what appears to be a trusted, established release.
A recent supply-chain attack has demonstrated that changes to dependency references can compromise an entire software ecosystem~\cite{paloalto_unit42_2025}.
Reproducibility can also be impacted when downstream software recipients assume that a tag will always point to the same commit, whereas in practice tags can be reassigned.
Yet despite these risks, no comprehensive study has focused on the phenomenon of tag alterations.

\paragraph{Problem statement}

Current software practices treat tag names as stable release identifiers~\cite{commonflow2010}, while Git semantics allow those identifiers to be reassigned or removed over time.
This creates a structural mismatch between what tools and users assume (a tag denotes a fixed reference) and what the underlying system guarantees (a tag is a mutable reference).
In practice, two builds that declare the same tag-based dependency may resolve to different source code at different times, undermining reproducibility and opening a supply-chain attack surface.
This risk is especially problematic for computational research workflows, where provenance and re-executability depend on stable, time-invariant references.

Despite the practical relevance of this mismatch, evidence at the ecosystem scale is still missing: we lack quantitative understanding of how often tag alterations occur, what kinds of alterations are most common, where they occur, and whether they already affect real package ecosystems.
Without such evidence, recommendations on tag-based dependency resolution remain largely ad hoc rather than empirically grounded.

\paragraph{Contributions}

The prevalence, characteristics, and impact of tag alterations in public code remain poorly studied.
A key reason is methodological: tag alterations are inherently difficult to study empirically at scale, because once a tag is altered, its previous reference no longer exists in the repository and cannot be directly compared with the post-alteration state.
To bridge this gap, we conduct the first large-scale quantitative and qualitative study of tag alterations across the public software ecosystem.
Using \SWH~\cite{swhcacm2018}---a comprehensive archive of publicly available source code---we analyze \TotalOriginsAnalyzedShort repositories over \TemporalSpan, spanning major forges including GitHub, GitLab, Bitbucket, Android, Codeberg, SourceHut, and others.
\SWH is critical for this study because it preserves successive snapshots of repository states with persistent identifiers~\cite{iso-swhid} (SWHIDs), allowing us to compare tag targets before and after an alteration even when the original forge state has already been overwritten.
Specifically, we address the research questions detailed below.

\begin{researchquestion}
  \label{rq:prevalence} How many tag alterations occur across major software forges and how do they evolve over time?
\end{researchquestion}
We show that tag alteration is not anecdotal: it appears at ecosystem scale, across major forges, and persists throughout the analyzed period.
This establishes tag mutability as an active, recurring practice rather than a rare historical artifact.

\begin{researchquestion}
  \label{rq:taxonomy} How are tags altered in software repositories?
\end{researchquestion}
We characterize how tags are altered by introducing a taxonomy that distinguishes between tag moves and deletions, as well as between moves that change the target of the tag to a new content from those that change the target to different metadata leaving the content similar.
We observe that alterations are heterogeneous, and that a meaningful fraction of moves retarget familiar tag names to states with different content.

\begin{researchquestion}
  \label{rq:correlation} Does repository popularity correlate with tag alterations?
\end{researchquestion}
We analyze alteration patterns against repository popularity indicators and package-ecosystem exposure.
The result is a nuanced pattern: low-popularity repositories dominate in absolute volume, but highly visible projects are also affected and cannot be treated as exempt.

\begin{researchquestion}
  \label{rq:threats} To what extent do tag alterations constitute a threat to software build reproducibility and supply-chain integrity?
\end{researchquestion}
We validate the threat model through both a software supply-chain attack case study and package-level cross-analysis.
Together, they show that tag alterations are not only a conceptual risk: they can produce concrete reproducibility drift and supply-chain integrity failures in real workflows.

\paragraph{Paper structure}
The remainder of the paper is organized as follows: \Cref{sec:background} introduces the Git and archival concepts, \Cref{sec:methodology} describes the pipeline and analyses, \Cref{sec:results} presents the findings, \Cref{sec:discussion} and \Cref{sec:threats-validity} discuss implications and validity threats, \Cref{sec:related} situates our work, and \Cref{sec:conclusion} concludes.

\section*{Data availability}
A complete reproducibility package for this paper is available on Zenodo~\cite{replication-package} and in \SWH (pending acceptance due to anonymization constraints), enabling reproducibility and future research.

 \section{Background}
\label{sec:background}

\paragraph{Git data model and tags}
\label{sec:background-git}

Git~\cite[§10.2, Git Objects]{ProGit2014} stores repository contents in a content-addressable object storage that uses cryptographic digests (SHA1, SHA256) as keys.
A Git repository, and hence the object storage, can contain different types of objects, most notably: blobs, trees, commits, and annotated tags.
Commits reference source trees and parent commits, while tags provide human-readable names for specific repository states, most commonly software releases.
In practice, tags are frequently used as stable version identifiers in build scripts, dependency declarations, and continuous integration (CI) workflows.

Git supports two main kinds of tags.
A \emph{lightweight tag} is a reference string (or simply ``ref'', in Git jargon) that points directly to an object, typically a commit.
An \emph{annotated tag} is a distinct Git object, present in the object storage, containing metadata such as a tagger name, timestamp, and message, and usually refers to a commit indirectly.
Despite this structural difference, both kinds of tags are mutable references: Git does not enforce that a tag name must always designate the same target over time.
Distinguishing them is relevant in our study because a tag alteration can be a transition from lightweight to annotated or vice versa.

The distinction between tag \emph{names} and the objects \emph{they point to} is central to our study.
Git objects are immutable once identified, but references to those objects are not.
As a result, a tag name may remain unchanged while its target changes.

\paragraph{Tag alteration in Git}
\label{sec:background-alteration}

Although users often treat tags as immutable release anchors, Git allows them to be deleted, recreated, or force-updated.
A maintainer can retarget an existing tag name to a different commit without changing the tag name itself.
These operations are neither necessarily malicious nor always undesirable.
They may result from release correction, repository cleanup, or migration.
However, once a tag has been published, official Git documentation explicitly discourages reusing the same name: the recommended approach is to issue a new version label (\emph{e.g.}, ``X.1'') rather than force-update ``X'', because two clones may both refer to ``version X'' while resolving to different commits~\cite{git_tag_retagging_docs}.
Linus Torvalds made the same point in 2007, arguing that silently overwriting an already-fetched tag would be a serious security problem and that maintainers who insist on force-updating a published tag must instead broadcast the change explicitly~\cite{torvalds_retagging_2007}.
These cautions matter because tag alterations become consequential for reproducibility and software supply-chain security when downstream users rely on tag names alone.
If a package recipe, CI configuration, or dependency declaration references a tag without ``pinning'' the underlying commit hash (e.g., by recording the hash digest the tag resolves to at the time it is first referenced), later tag alteration may silently change the retrieved source code.

\begin{figure}
  \centering
  \includegraphics[width=\linewidth]{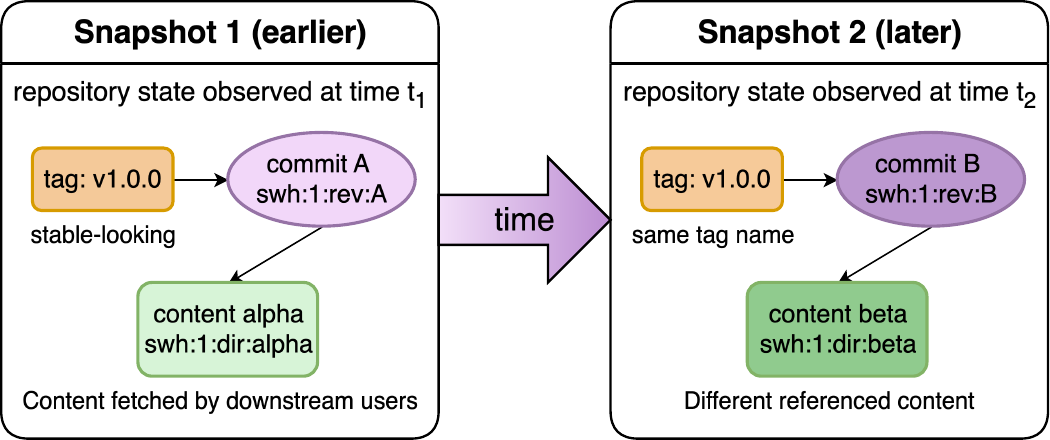}
  \caption{Tag \emph{move} observed between two snapshots of the same Git repository: tag name (\texttt{v1.0.0}) resolves to different commits before/after the alteration.}
  \label{fig:tag-alteration}
\end{figure} 

This paper focuses on two observable forms of tag alteration: (1) \emph{moves}, where a tag is observed in two snapshots but resolves to different targets, as illustrated in \Cref{fig:tag-alteration}, or where a tag disappears but is recreated later, and (2) \emph{deletions}, where a previously observed tag disappears from one snapshot to another.

\paragraph{\SWH}
\label{sec:background-swh}

\SWH~\cite{swhcacm2018} is a universal archive of publicly available source code.
For Git repositories, it regularly visits \emph{origins}---repository URLs---and records their observed states as \emph{snapshots}.
Each snapshot captures the set of visible references---including tags---together with the archived objects they designate.
\SWH periodically crawls each origin to take snapshots, then stores the objects in a global Merkle Directed Acyclic Graph (DAG).
\SWH assigns Git-compatible persistent identifiers (SWHIDs~\cite{iso-swhid}) to archived objects such as revision objects (\textit{e.g.} commits for Git), directory objects, releases, and snapshots, which allows us to distinguish changes in commit identity (revision SWHID) from changes in source tree content.

The remainder of the paper builds on this archival view.
\Cref{sec:methodology} explains how we compare consecutive \SWH snapshots to detect and classify tag alterations across large numbers of repositories.
 \section{Methodology}
\label{sec:methodology}

To answer the stated research questions, we follow an empirical methodology consisting of four phases: (1) we collect the largest existing corpus of periodically crawled public repositories for analysis (required for \Cref{rq:prevalence}); (2) we devise an experimental protocol to detect tag alterations (addressing \Cref{rq:prevalence} and \Cref{rq:correlation}); (3) we create a taxonomy to categorize tag alterations and understand the technical scope of the changes (\Cref{rq:taxonomy}); and (4) we assess reproducibility and security implications through case studies and threat analysis (\Cref{rq:threats}).

\begin{figure*}
    \centering
    \includegraphics[width=0.95\textwidth]{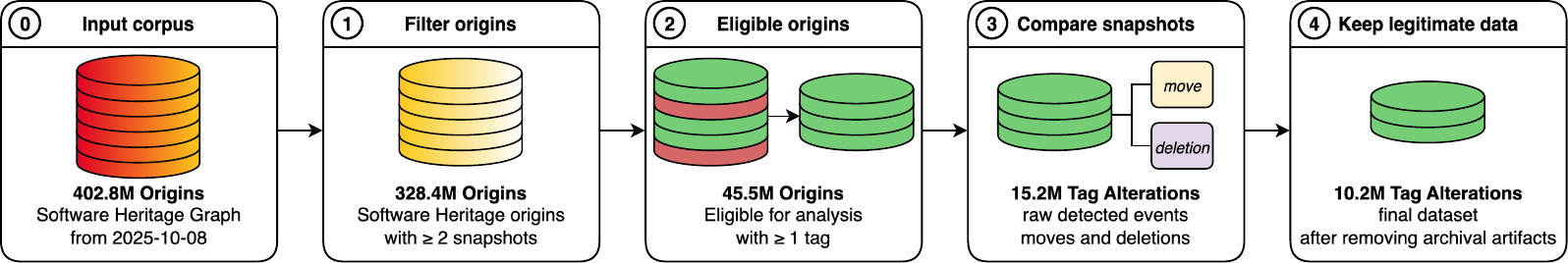}
    \caption{Overview of the workflow pipeline. Step~0 fetches all available origins from \SWH. Step~1 identifies candidate origins. Step~2 selects eligible repositories. Step~3 detects alterations by comparing consecutive snapshots and categorizes moves and deletions. Step~4 removes the archival artifact to keep only true alterations.}
    \label{fig:workflow-pipeline}
\end{figure*}

\subsection{Data collection and repository selection}
\label{sec:data-source}

We leverage \SWH~\cite{swhcacm2018} to analyze repositories archived from major forges including GitHub, GitLab, Bitbucket, and others.
Our study uses the \SWH graph snapshot from \DatasetDate~\cite{swh-msr-2020-challenge}, which preserves historical repository states even after tags are altered or deleted, enabling our analysis.

From the complete \SWH graph---\Cref{fig:workflow-pipeline}, step 0---, we apply a series of filters to identify repositories suitable for tag alteration analysis.
We require repositories with at least \MinSnapshotsRequired snapshots to enable temporal comparison---represented in \Cref{fig:workflow-pipeline}, step 1---, which filters out \OriginsFilteredOutShort origins captured only once, leaving \OriginsWithMultipleSnapshotsShort origins with multiple snapshots.
Due to the Git scope of this work, we filter out non-Git origins as well as incomplete ones (called ``partial'' in SWH terminology).
We then select only repositories containing at least one tag across their histories---\Cref{fig:workflow-pipeline}, step 2---, yielding \OriginsWithTagsShort relevant repositories.
Finally, we exclude alterations before \CutoffDate, due to a known \SWH archival artifact that produced spurious alterations, which we have confirmed with the archive operators (\NonLegitAlterationsShort false positives)---\Cref{fig:workflow-pipeline} step 4.
All alterations after this date are considered legitimate.

To answer \Cref{rq:correlation}, we group repositories by popularity using GitHub ``stars'' as a metric~\cite{borges-2016-github-stars}.
Although the number of stars can be artificially inflated by various means~\cite{DBLP:journals/corr/abs-2412-13459}, it remains a relevant metric in our case: a large number of stars increases repository visibility, making the repository more worthy of scrutiny.
Due to the lack of popularity metrics on other development platforms, analyses that depend on popularity focus on GitHub repositories, while other analyses are conducted on all repositories archived by \SWH, regardless of the platform.

\subsection{Tag alteration detection}
\label{sec:detection-algorithm}

We detect tag alterations by comparing consecutive snapshots of each repository---\Cref{fig:workflow-pipeline}, step 3.
For each \SWH origin with at least two visits, we extract all snapshots ordered chronologically by timestamp.
For each snapshot, we enumerate all branches whose names match the pattern \texttt{*/tags/*}; this is because Git stores tags as special refs in the \texttt{refs/tags/} namespace.
We classify each tag as either lightweight (a reference pointing directly to a commit) or annotated (a reference pointing to a release object, which in turn references a commit).
We then resolve the target commit for each tag: lightweight tags directly reference a commit, while annotated tags require traversing the release object to reach the target commit.

For each pair of consecutive snapshots $(S_i, S_{i+1})$, we identify tags present in $S_i$ and compare them with $S_{i+1}$ to detect two primary types of events: (1)~a \emph{move} occurs when a tag name exists in both snapshots, but points to a different object (identified by a SWHID), and (2)~a \emph{deletion} occurs when a tag exists in $S_i$ but not in $S_{i+1}$.
For each detected alteration, we record metadata including repository URL, tag name, and tag type (lightweight vs.~annotated).

\subsection{Tag alteration categorization}
\label{sec:classification}

\begin{table}
\centering
\small
\caption{Distribution of \emph{move} categories.}
\label{tab:diff-categories}
\begin{tabular}{lrr}
\toprule
\textbf{Category} & \textbf{Count} & \textbf{Share (\%)} \\
\midrule
Content changes & \num{426595} & \num{68.15} \\
Commit metadata diffs  & \num{188974} & \num{30.19} \\
Release-level changes   & \num{10439}  & \num{1.67}  \\
\midrule
\emph{Total}           & \num{626008} & \num{100.00} \\
\bottomrule
\end{tabular}
\end{table}

To understand the scope and potential impact of tag alterations (addressing \Cref{rq:taxonomy}), we develop a classification taxonomy, shown in \Cref{tab:diff-categories} and \Cref{tab:diff-fields}, by systematically characterizing which part undergoes modification.
Because a Git tag ultimately resolves to a source tree, an alteration can modify different layers of the Git object hierarchy.
We categorize the nature of the change by comparing the old and new SWHIDs at three distinct levels:
\begin{itemize}
\item \textbf{Content (source tree) changes:} The underlying source code tree (directory SWHID) is modified, meaning files were added, removed, or edited.
  This represents a true change to the software's content.
\item \textbf{Commit metadata changes:} The source tree remains identical (no source code changed), but the revision SWHID (commit identity) changes.
  This typically indicates a rewritten commit history where metadata such as the commit timestamp, author, or commit message was altered.
\item \textbf{Tag/Release-level changes:} Both the source tree and commit identities remain identical, but the tag reference layer changes.
  This includes transitions between lightweight and annotated tags, as well as deletion--recreation patterns that preserve the same commit and source tree targets.
  We also check annotated-release metadata fields (message, author, and author timestamp) separately; in our dataset, we did not capture any occurrence of an alteration modifying those fields.
\end{itemize}

\noindent
Additionally, we track temporal patterns, distinguishing between a standard \textbf{move} (the tag target is updated directly between consecutive snapshots) and a \textbf{recreation} (a tag is deleted in one snapshot and reappears pointing to, either the same commit, or a different one in a later snapshot).
This hierarchical classification allows us to precisely distinguish benign metadata (\emph{e.g.}, commit message update) updates from potentially malicious or disruptive source code modifications.

\subsection{Popularity measurement}
\label{sec:statistical-analysis}

To address \cref{rq:correlation}, we collect an additional repository characteristic and perform basic correlation analysis.
For GitHub repositories, we collect star counts as a proxy.

We categorize repositories into star ranges: \StarRangeZeroOne, \StarRangeTwoTen, \StarRangeElevenHundred, \StarRangeHundredFiveHundred, and \StarRangeFiveHundredPlus stars.
The 0--1 range is grouped together to account for the fact that repository owners can star their own repository; it is hence safer to use 2 as threshold for evidence of ``non zero'' popularity.

To assess the relationship between popularity and tag alteration, we first analyze the proportion of tag alterations across the selected star ranges, together with an Unknown category.
The Unknown category includes repositories hosted on non-GitHub forges and GitHub repositories for which star-count information could not be retrieved.
We then restrict the analysis to repositories with known star counts and compute, for each star range, the proportion of repositories that exhibit at least one tag alteration.
This measure indicates in which popularity range repositories are more likely to have altered at least one tag.

\subsection{Case studies and threat assessment}
\label{sec:threat-modeling}

To evaluate the security implications of tag mutability (\cref{rq:threats}), we present an in-depth case study of a known software supply chain attack on the \texttt{\footnotesize{tj-actions/changed-files}} repository~\cite{paloalto_unit42_2025}.
Specifically, we examine a fork of that repository, which was captured in our dataset before and after the attack.
This case study provides evidence of how attackers can leverage tag alterations to facilitate malicious code injection.

Furthermore, we assess the threat that tag alterations pose to build reproducibility by cross-referencing our dataset with the Nixpkgs ecosystem, specifically using the Nixpkgs master branch on the date of the experiment~\footnote{\url{https://github.com/NixOS/nixpkgs/archive/08ffd21f7c8f90e15216bc329855b66ffea8526d.tar.gz}}.
First, we identify the subset of Nix packages that depend on repositories with a history of tag alterations.
Second, we isolate the specific Nix packages that directly pin their dependencies to a historically altered tag.
For this vulnerable subset, we systematically compare the cryptographic hash of the source tarball expected by the Nix package against the actual hash of the tag's target code post-alteration.
This comparative analysis allows us to identify concrete packages whose source-based build reproducibility is threatened by tag alterations.

Nixpkgs comes with a binary cache where package sources are also stored.
This means that rebuilding a package with an altered tag may still succeed if the source tarball is available in the cache.
If the cache becomes unavailable, Nix's integrity checks will fail, meaning that the package will not build, rather than building with the altered source code.
This avoids silent reproducibility drift or the introduction of malicious changes, but still constitutes a supply-chain availability failure.

 \section{Results}
\label{sec:results}

This section answers the research questions stated in \Cref{sec:intro}.
Following the methodology presented in \Cref{sec:methodology}, we first measure tag alterations and their evolution over time, then characterize their observable forms through a structured taxonomy, next examine whether repository popularity is associated with alteration behavior, and finally assess the implications for reproducibility and software supply-chain integrity.

\subsection{RQ1: Quantity and evolution over time of tag alterations}
\label{sec:results-rq1}

A first question is whether tag alterations are anecdotal exceptions or a widespread phenomenon at ecosystem scale.
To answer this, we analyzed successive snapshots of repositories archived in \SWH and identified \TotalAlterations tag alterations affecting \AlteredRepositories unique repositories (\AlteredRepositoriesPercent of repositories with at least two snapshots and one tag).
This result establishes that tag mutability is not a marginal corner case: alterations occur at a scale of more than ten million events across nearly two hundred thousand repositories.

Among these alterations, \LegitMoves are moves and \Deletions are deletions.
Deletions account for the overwhelming majority of observed events, while moves represent a smaller but important subset because the tag name remains present while silently changing its target.
This distinction matters for downstream users: a deleted tag is visible as a failure or absence, whereas a moved tag preserves the \emph{appearance} of continuity while changing the referenced object.

While the proportion of repositories with tag alterations is small, the number of alterations per repository is rather high: \AlteredTagPerRepository on average (median: \AlteredTagPerRepositoryMedian).
The sheer number of tag deletions contributes significantly to this high average and can partially be explained by edge cases such as the repository \texttt{intellij-community} from JetBrains.
This repository had to delete tens of thousands of tags because of performance issues with Git.\footnote{https://blog.jetbrains.com/platform/2018/11/cleaning-up-tags-on-github/}
This repository alone, together with \num{28} of its forks, accounts for \num{742 919} tag deletions.

\begin{figure}
    \centering
    \includegraphics[width=\linewidth]{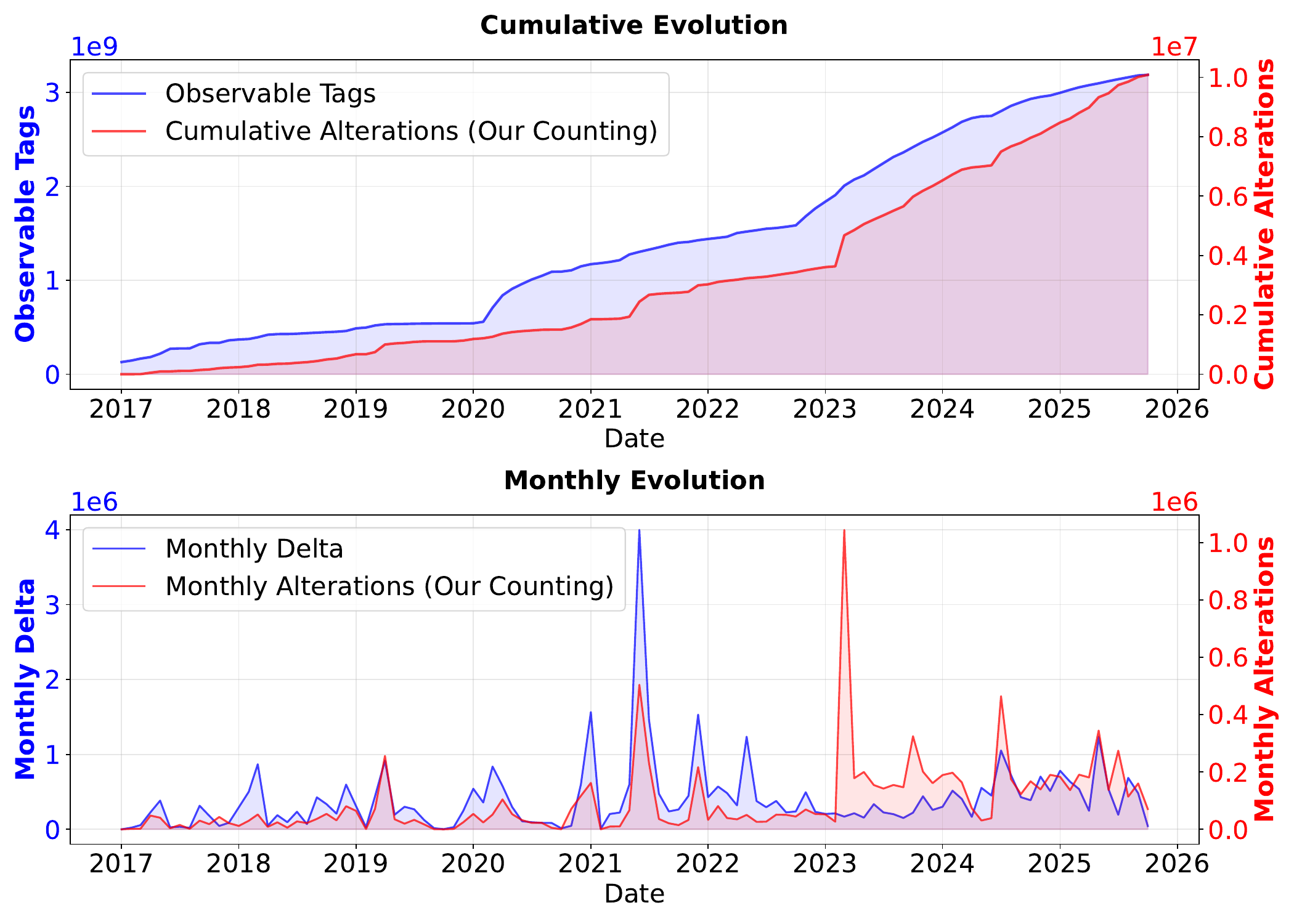}
    \caption{Evolution of the number of observable tags in \SWH vs.\ number of tag alterations (note the different scales)}
    \label{fig:temporal-evolution}
\end{figure}

The temporal analysis displayed in \Cref{fig:temporal-evolution} shows that tag alterations are not confined to a short-lived historical period.
Over the observation window, the cumulative number of alterations grows continuously, reaching \TotalAlterations by the end of the analyzed period.
The monthly aggregation used for plotting, which spans from \CutoffDate to \DatasetDate, shows persistent accumulation rather than isolated bursts.
This sustained growth indicates that tag alteration is a recurring maintenance or release-management behavior in the ecosystem, not an artifact of a few exceptional incidents.

If tag alterations were mostly caused by early ecosystem instability or a few legacy forges, one would expect the curve to flatten over time.
Instead, the cumulative increase suggests that the practice remains active as the ecosystem grows.
This interpretation is consistent with the monthly (non-cumulative) view in the bottom plot of \Cref{fig:temporal-evolution}: although a few months exhibit pronounced spikes, alterations are observed throughout the timeline, indicating a persistent baseline of activity rather than a one-off historical wave.
Because our method only captures alterations visible between \SWH snapshots, this trajectory should be interpreted as a lower bound on observable occurrences, which further strengthens the conclusion that tag mutability is widespread.

These results answer RQ1: tag alterations occur frequently and persist throughout the \TemporalSpan period.
The core implication is that any reproducibility or supply-chain model that treats tags as \emph{de facto} immutable is misaligned with observed practice.

\subsection{RQ2: How are tags altered?}
\label{sec:results-rq2}

The second question concerns how tags are altered.
Our dataset allows us to recover the \emph{observable behavior} of altered tags.
Rather than claiming to infer motives, we derive a taxonomy from the concrete transition patterns visible between snapshots.

Our first result is that the dominant observable behavior is \emph{deletion}.
Out of the \TotalAlterations alterations, \Deletions (\DeletionsPercent) correspond to tags disappearing between consecutive observations.
This shows that the most common alteration pattern is not silent retargeting but removal.
Deletions primarily threaten software availability and rebuildability: historical references may cease to resolve, causing previously reproducible workflows that rely on them to fail.

\begin{table}
\centering
\small
\caption{Field-level distribution within each \emph{move} category.}
\label{tab:diff-fields}
\begin{tabular}{llSS}
\toprule
& \textbf{Field} & \textbf{Count} & \textbf{Share (\%)} \\
\midrule
\multicolumn{4}{l}{\textit{Content changes (denominator: content diffs, $N=$ \num{426595})}} \\
& file added     & \num{219117} & \num{51.36} \\
& file deleted   & \num{217789} & \num{51.05} \\
& file modified  & \num{333283} & \num{78.13} \\
& file renamed   & \num{63496}  & \num{14.88} \\
\addlinespace[0.4ex]
\multicolumn{4}{l}{\textit{Commit metadata only ($N=$ \num{188974})}} \\
\addlinespace[0.4ex]
\multicolumn{4}{l}{\textit{Release changes (denominator: release diffs, $N=$ \num{10439})}} \\
& Deletion $\rightarrow$ Recreation (temporary missing) & \num{160} & \num{1.53} \\
& Lightweight $\rightarrow$ Annotated        & \num{7555}     & \num{72.37}  \\
& Annotated $\rightarrow$ Lightweight         & \num{2724}    & \num{26.09}      \\
\bottomrule
\end{tabular}
\end{table}

The second major observable behavior is \emph{move}, where the tag name persists but points to a different target.
We identify \LegitMoves such move events, including both direct moves and deletion--re\-creation patterns that recreate the tag with a different target.
Although numerically smaller than deletions, moves are more security-sensitive because they preserve the familiar tag identifier while changing the referenced state.
In other words, moves are the class of alteration most directly capable of violating the intuitive assumption that a release tag designates a stable version and is immutable.

Among move events, content-changing moves are the dominant class (\PercentMovesContentChange).
We then refine the remaining moves into two distinct sub-types of metadata-only moves, consistent with \Cref{sec:classification}.
We start by discussing the metadata-only moves before coming back to the content-changing ones.
First, \textit{commit-metadata-only} moves preserve the content but change the commit object; this corresponds to \RevisionDiff cases (\RevisionDiffPercent) in \Cref{tab:diff-categories}, with differences distributed across commit metadata.
Second, \textit{release-level-only} changes preserve both commit and content targets and account for \ReleaseDiff cases (\ReleaseDiffPercent).
Within this release-level subset, we observe no annotated-release metadata differences (message, author, author timestamp); the observed cases are structural reference changes, namely lightweight $\leftrightarrow$ annotated transitions and temporary deletion--recreation with the same target.

A moved tag could be a harmless correction---for example, converting a lightweight tag to an annotated tag without changing commit or source tree targets, or temporarily deleting and recreating the same tag target.
But in the majority of observed move events, the underlying content actually changes.
Therefore, the typical move in our dataset is not only administrative; it alters the code that downstream users obtain when resolving the tag.

The source-tree-level breakdown further characterizes the nature of these content-changing moves.
As shown in \Cref{tab:diff-fields}, among source tree diffs ($N=\num{426595}$), file modifications are the most frequent signal (\num{333283}, \num{78.13}\%), while file additions (\num{219117}, \num{51.36}\%) and deletions (\num{217789}, \num{51.05}\%) are also common; strict renames are less frequent (\num{63496}, \num{14.88}\%).
Because these proportions are not mutually exclusive, they do not directly separate cosmetic from substantive edits.
Still, they provide a conservative lower bound: at least \num{51.36}\% of source tree diffs include a file addition or deletion, indicating that structurally non-trivial changes are common among content-changing moves.

Tag deletions primarily affect availability; metadata-only moves challenge bit-level identity, but may preserve source content; finally, content-changing moves directly threaten the integrity of tag-based dependency resolution.

\subsection{RQ3: Correlation with repository popularity}
\label{sec:results-rq3}

To examine whether repository popularity correlates with alteration behavior, we use GitHub star counts as a proxy for repository visibility and group repositories into the ranges \StarRangeZeroOne, \StarRangeTwoTen, \StarRangeElevenHundred, \StarRangeHundredFiveHundred, and \StarRangeFiveHundredPlus stars.

\begin{figure}[!h]
    \centering
    \includegraphics[width=\linewidth]{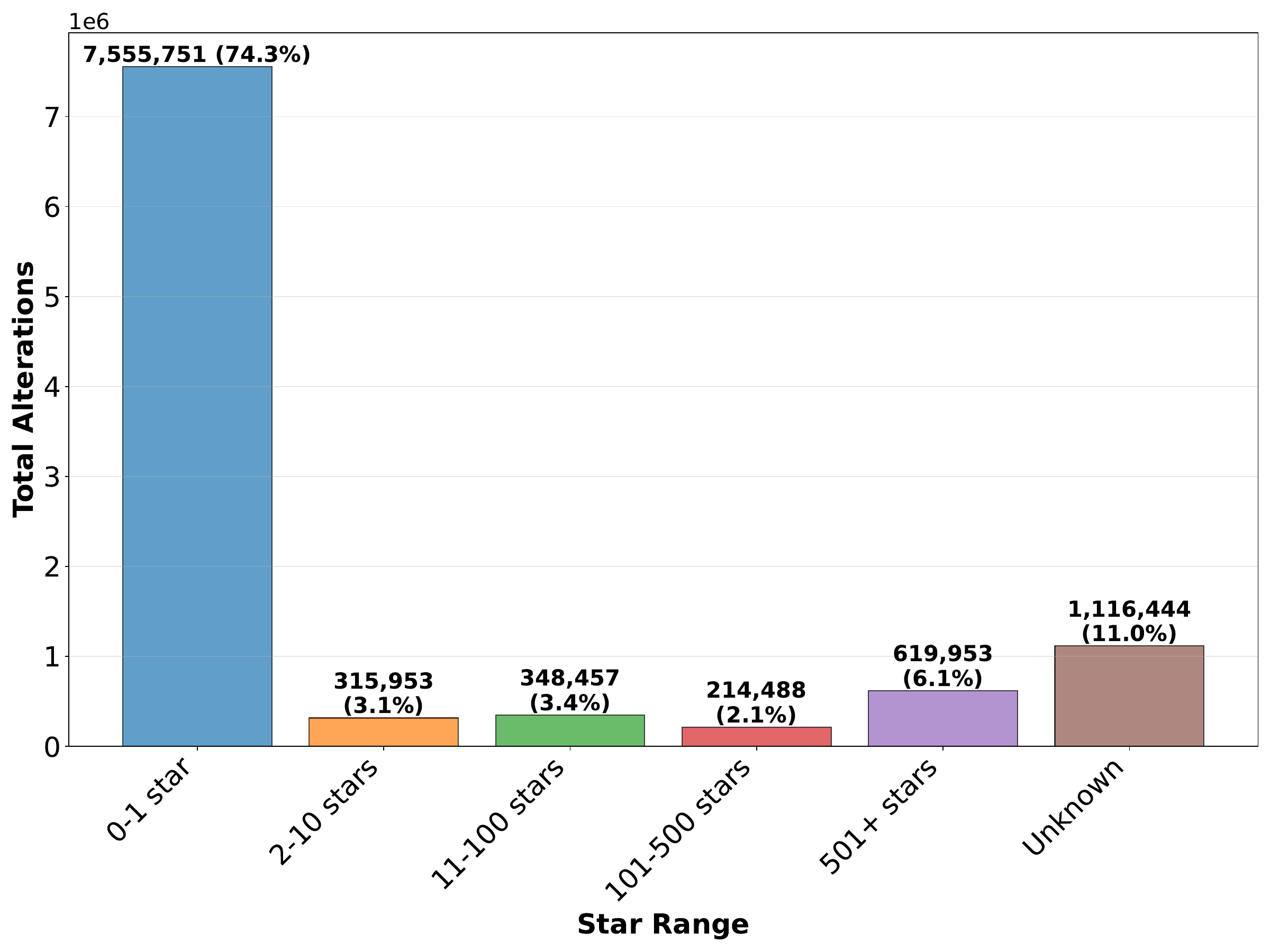}
    \caption{Total number of alterations by repository star count}
    \label{fig:popularity_star_count}
\end{figure}

In absolute terms, tag alterations are overwhelmingly concentrated in the least popular repositories.
Repositories with only \StarRangeZeroOne stars account for \StarZeroOneAlterations alterations, i.e., \StarZeroOneAlterationsPercent of all analyzed alterations, as shown in \Cref{fig:popularity_star_count}, and represent \StarZeroOneOrigins repositories, or \StarZeroOneOriginsPercent of all affected repositories.
This concentration shows that, at ecosystem scale, most observed tag instability comes from the long tail of low-visibility repositories rather than from a small number of highly visible projects.

\begin{figure}[!h]
    \centering
    \includegraphics[width=\linewidth]{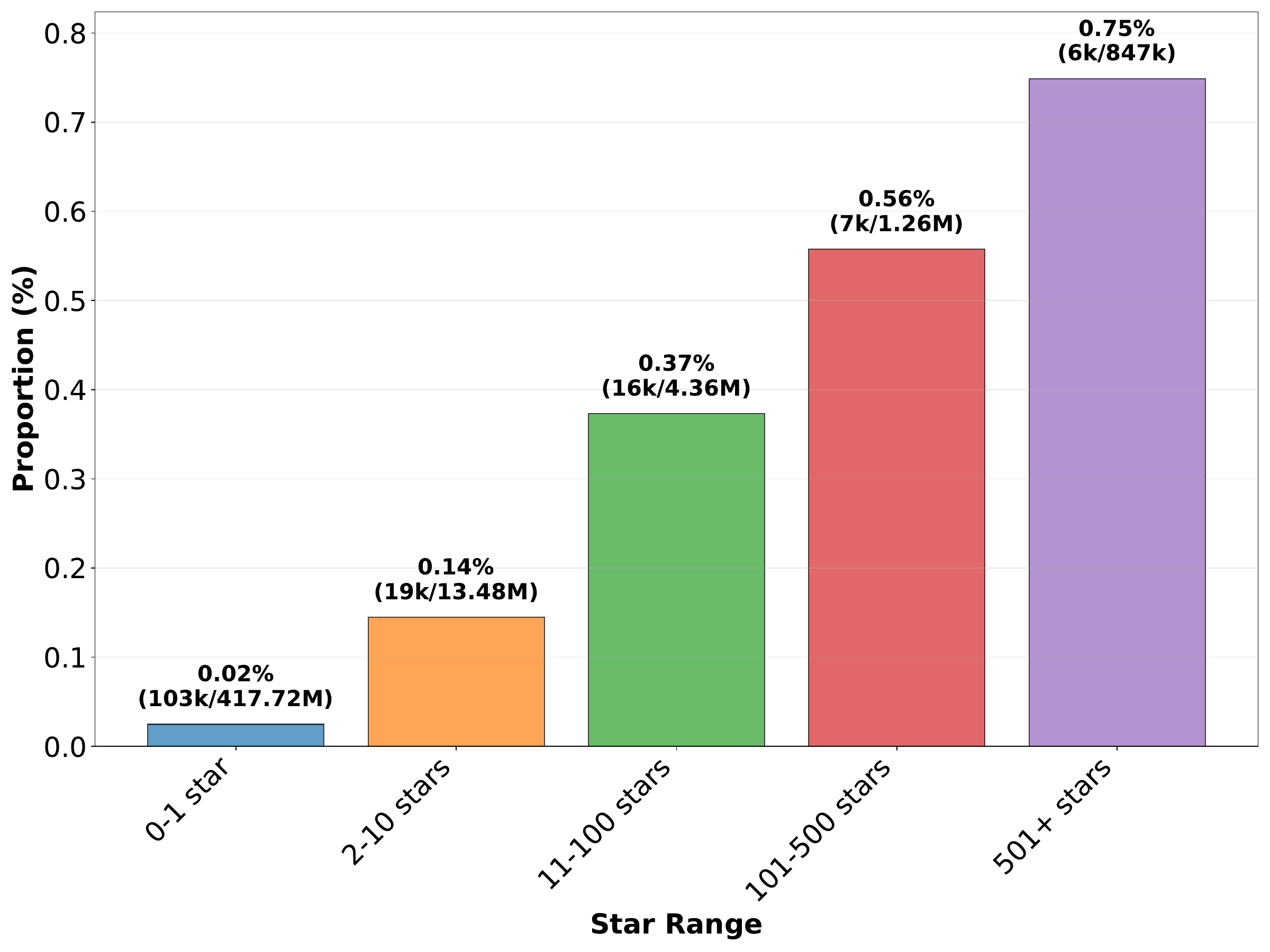}
    \caption{Proportion of repositories with $\ge 1$ alteration by star count}
    \label{fig:popularity_proportion}
\end{figure}

\Cref{fig:popularity_proportion} provides a perspective on the proportion of repositories with at least one observed alteration in each star range.
This proportion increases steadily with popularity: approximately \num{0.0248}\% for \StarRangeZeroOne, \num{0.1447}\% for \StarRangeTwoTen, \num{0.3732}\% for \StarRangeElevenHundred, \num{0.5576}\% for \StarRangeHundredFiveHundred, and \num{0.7488}\% for \StarRangeFiveHundredPlus.
Thus, although the total volume of alterations is dominated by low-star repositories, the likelihood that a repository exhibits at least one alteration increases with popularity.

Taken together, these results show that popularity correlates with alteration behavior, but not in a simple monotonic way if one considers only raw counts.
Low-popularity repositories dominate the phenomenon in absolute volume because they are extremely numerous and collectively account for most altered repositories.
By contrast, highly popular repositories are more likely to exhibit at least one alteration, and when they do, they can accumulate more alterations per affected repository--- among repositories with at least one tag alteration, the average amount of tag alteration per repository is \num{73} for \StarRangeZeroOne star and \num{98} for \StarRangeFiveHundredPlus stars.

Overall, we can see that repository popularity is meaningfully associated with tag alteration behavior.
The long tail of low-star repositories explains most alterations in aggregate, whereas highly popular repositories show the highest proportion of repositories with at least one alteration and the strongest per-repository intensity.
Tag mutability should therefore not be framed solely as a problem of obscure repositories or solely as a problem of prominent ones; rather, popularity changes how the phenomenon manifests.

\subsection{RQ4: Threats to reproducibility and software supply-chain integrity}
\label{sec:results-rq4}

The final research question is whether tag alterations pose a practical threat to build reproducibility and software supply-chain security.
Our results show that the threat is real for two independent reasons: first, many observed moves change the actual content behind a stable tag name; second, the phenomenon already intersects with real package-management metadata.

Among \LegitMoves move events, \MovesWithContentChange (\PercentMovesContentChange) change the referenced source tree content.
This is the exact failure mode that undermines tag-based pinning.
When a build, package recipe, or CI workflow resolves a dependency by tag name alone, a later content-changing move can silently substitute new code while preserving the same human-readable version identifier (e.g., v1.0).
In other words, the same dependency specification no longer targets the same source tree, and the trust attached to a release label is undermined.

This threat model is not hypothetical.
Our dataset captures a concrete software supply-chain attack case in the repository \texttt{\footnotesize{zendesk/changed-files}}, a fork or mirror of \texttt{\footnotesize{tj-actions/changed-files}}.
On March 14, 2025, we observe \num{346} tags being moved so that they all resolve to the same malicious commit, identified in \SWH as \texttt{\footnotesize{swh:1:rev:0e58ed8671d6b60d0890c21b07f8835ace038e67}}.
This pattern is highly atypical from a release-engineering perspective: a large set of pre-existing version labels is simultaneously retargeted to one identical commit.
According to incident reports~\cite{paloalto_unit42_2025}, the retargeted commit injected code that dumped CI runner memory into workflow logs, exposing environment variables and secrets---API tokens, credentials, and other sensitive values---of any project whose pipeline invoked the action.
Because \texttt{\footnotesize{tj-actions/changed-files}} is a widely reused GitHub Action, the tag retargeting did not affect a single consumer in isolation: every workflow that resolved the action by an existing version tag would execute the same malicious payload on its next run, even when developers had pinned a specific tag such as \texttt{v39} or \texttt{v47}.
For affected teams, the consequences were therefore immediate and operational: audit workflow logs for leaked secrets, rotate compromised credentials, and update workflow definitions to reference known-good commit hashes rather than mutable tag names.

The temporal profile of these tag \emph{moves} further illustrates the reproducibility risk.
Before being altered, the affected tags had remained stable for at least \num{148} days in \num{340} cases, with a few shorter lower-bound intervals of \num{6}, \num{37}, \num{81}, and \num{127} days.
These values show that the tags were not ephemeral placeholders changed immediately after creation; many had existed long enough to plausibly be consumed, cached, mirrored, or trusted as stable release references.
Once retargeted, any downstream workflow that fetched the same tag name again would no longer obtain the same source tree.

The second line of evidence comes from the cross-analysis with Nixpkgs.
We identify \NixPkgsTracked packages whose source references point to a repository in our dataset that altered at least one tag.
Narrowing this set, we find \NixPkgsAffected distinct altered tags explicitly referenced in Nix packages.
Most importantly, we identify \NixPkgsHashMismatch packages that failed to build when disabling the NixOS cache: \num{4} failed because the tag was deleted and the process could not fetch the source, and \num{3} failed because the fetched content no longer matches the previously stored hash.
The detailed enumeration of packages with a build error outcome is provided in \Cref{tab:nixpkgs} in the Appendix.
This evidence shows that tag alterations are not only theoretically capable of breaking the reproducibility of software builds; they have done so in real packages in a major Linux distribution.
For the \num{25} other packages, there is no hash mismatch, meaning that the Nix package definitions were created (or updated, although that is less likely) after the alteration.
The command-level evidence in \Cref{fig:nix-build-hash-mismatch} illustrates a rebuild failure caused by source retargeting.

Taken together, the tag-retargeting case study, the Nix cross-analysis, and the software supply-chain attack answer RQ4 affirmatively.
Tag alterations constitute a practical threat to both software build reproducibility and supply-chain integrity.

\begin{figure}[h]
\begin{lstlisting}[language=bash,basicstyle=\ttfamily\footnotesize,columns=fullflexible,breaklines=true]
$ nix build github:NixOS/nixpkgs/08ffd21[...]#libv3270.src \
--rebuild --no-link --impure -L
source> structuredAttrs is enabled
source>
source> trying https://[...]/refs/tags/5.5.0.tar.gz
[...]
FAIL github:NixOS/nixpkgs/08ffd21[...]#libv3270.src
error: hash mismatch in fixed-output derivation '/nix/[...]':
         specified: sha256-Cn/to1/[...]=
            got:    sha256-tW3CPga[...]=
\end{lstlisting}
\caption{Nix rebuild attempt for \texttt{libv3270.src} showing a fixed-output hash mismatch after tag retargeting.}
\label{fig:nix-build-hash-mismatch}
\end{figure}

 \section{Discussion}
\label{sec:discussion}

Our results show that tag mutability is not only a theoretical property of Git: it is used in practice.
Nor can it be dismissed as a quantitatively marginal phenomenon: at the scale of public code archives, we observe tag alterations affecting a substantial number of repositories, with a non-trivial subset of move events changing the source tree content.
From a reproducibility perspective, the main problem is therefore not only that tags \emph{can} change, but that they may be relied upon as if they identified a fixed source state.

\paragraph{Tag mutability as a reproducibility problem}

At the level of Git semantics, tag mutability is allowed: tags can be deleted, recreated, or force-updated.
At the level of ecosystem practice, however, tags function as release identifiers, version shortcuts, or human-readable dependency anchors.
Our results show that these two layers do not align cleanly.
When a tag is reused to point to new content, the identifier remains stable while the referenced source tree changes.
In other words, the same dependency specification can resolve to different code over time.

Our taxonomy further refines this point.
Not all tag alterations have the same impact.
Deletions are numerous, but they are comparatively visible to downstream consumers because the reference disappears and retrieval may fail explicitly.
Moves are less frequent, but more consequential for integrity because the tag name remains available while silently changing meaning.
Likewise, not all moves change software content: some only affect commit- or release-level metadata.
However, the fact that a majority of moves modify the source tree content demonstrates that tag-based dependency resolution cannot, by itself, guarantee the stability of the retrieved source artifact.

The Nixpkgs cross-analysis strengthens this argument by connecting our observations to a package ecosystem.
We identify a subset of packages with actual hash mismatches between the previously expected artifact and the content retrieved at analysis time
(plus some where the tag has been removed).
This establishes an evidence chain from tag alteration to explicit dependency reference to observable reproducibility failure.
Tag mutability is not only a conceptual threat to reproducibility; it has already materialized as divergence between expected and obtained source artifacts.

\paragraph{Implications for software supply-chain integrity}

The software supply-chain attack case in our dataset provides an example of the mechanism identified throughout this paper: a familiar tag can keep its name while silently changing the code it resolves to.
The implication for supply-chain integrity is therefore semantic, not just technical.
Once a mutable reference is socially interpreted as a stable release handle, downstream users and tools may trust continuity that no longer exists.

The risk created by tag mutability is not restricted to traditional package managers.
Convenience solutions, such as relying on tags or floating references, conflict with strong reproducibility guarantees unless they are paired with immutable identifiers or independent integrity verification (as Nixpkgs does).
Our results therefore support a general design principle: human-friendly release names may remain useful for discovery and communication, but build-relevant resolution should ultimately be grounded in cryptographic identifiers.

\paragraph{Ecosystem ambiguity and the case of GitHub Actions}

The GitHub Actions ecosystem provides a useful example of this tension.
In that ecosystem, the mutability of tags is not entirely hidden: some official guidance has encouraged maintainers to move major-version tags~\cite{github_actions_versioning_recommendations}, while security guidance for consumers emphasizes pinning third-party actions to immutable commit hashes~\cite{github_secure_use_third_party_actions}.
This does not make the ecosystem incoherent so much as it reveals that tags play two conflicting roles at once.
For maintainers, a mutable tag can act as a moving compatibility label; for consumers, the same tag may be interpreted as a stable release anchor.
The problem is not simply that developers are unaware that tags can move.
Rather, different communities and tools assign different normative meanings to the same reference type.
Reproducibility failures can emerge when these meanings are conflated.

This point also helps position our contribution relative to common engineering practice.
Reproducibility research is concerned not only with what mechanisms are technically possible, but with whether they produce observable instability in real ecosystems.
Our results show that they do.
Tag alteration is measurable at large scale, content-changing moves are common enough to matter, and the phenomenon intersects with real package metadata and a concrete attack pattern.
The relevant question is therefore not whether tag mutability exists, but how often it occurs, what forms it takes, and under what conditions it undermines downstream assumptions.

Some communities have partially integrated the idea that certain tags behave as moving labels.
At the same time, the same ecosystems also rely on tag-like references as convenient dependency handles, while separately recommending cryptographic pinning when stronger integrity guarantees are required.
The coexistence of these two models is precisely what makes the practice fragile: a reference that is socially treated as both a release name and a mutable pointer is liable to be interpreted differently by maintainers, users, or tools.

\paragraph{Implications for tools, platforms, and package ecosystems}

These findings have several implications for tool builders and platform operators.
First, build tools and package managers that target reproducibility should not treat tag names as sufficient dependency anchors on their own.
The practical compromise is to record both the human-readable tag name and the cryptographic commit or content hash it resolved to at the time of reference, as Nixpkgs does when storing expected source hashes.
This preserves the convenience of tag-based version selection while pinning the actual artifact identity.
These recommendations are not merely precautionary: ecosystem experience shows that re-tagging has repeatedly caused concrete failures.
In Go modules, a 2025 incident on \texttt{google/go-containerregistry} illustrates a failure mode: after \texttt{v0.20.4} was deleted and recreated with additional commits, consumers encountered a checksum mismatch, forcing the project to publish \texttt{v0.20.5} and document \texttt{v0.20.4} as unusable as a Go module~\cite{go_containerregistry_retagging_issue}.
Second, code-hosting platforms could improve transparency by exposing native audit trails for tag creation, movement, and deletion, making reference mutability easier to monitor and reason about.

An example of platform-level response to the tag mutability problem is GitHub's introduction of immutable releases in August 2025~\cite{github_immutable_releases}.
It is important to note that this feature applies specifically to GitHub Releases---releases created through the GitHub web interface or the \texttt{gh} CLI tool---rather than to git tags in general.
Once enabled (the feature is opt-in and disabled by default), GitHub Releases become immutable: the associated git tag cannot be moved or deleted, and release assets cannot be modified.

This implementation illustrates both the potential and the limitations of platform-level immutability enforcement.
On the one hand, it directly addresses the tag alteration patterns documented in this paper by preventing the most common forms of tag manipulation for releases that opt into protection.
On the other hand, several factors limit its scope: the feature must be manually enabled for each repository, and it applies only to GitHub Releases rather than to arbitrary git tags.
More broadly, because immutability is enforced at the platform level rather than in the git protocol itself, protections do not extend to other forges or to direct git operations outside GitHub's infrastructure.

More broadly, our results suggest that tag mutability should be understood as a socio-technical reproducibility problem rather than a version-control feature.
The issue does not arise solely from Git's permissive reference model, nor solely from downstream misuse.
It emerges from the interaction among flexible developer workflows, platform conventions, security advice, package metadata, and user expectations.

\section{Threats to validity}
\label{sec:threats-validity}

\paragraph{Construct validity.}
Our definition of a tag alteration relies on differences observed through \SWH snapshots and SWHID comparisons.
This provides a scalable way to detect changes to releases, commits, and source trees, but it does not directly expose the intention behind a change.
In particular, we can determine that a tag was deleted, moved, or otherwise altered between observed states, yet, based on archival evidence alone, we cannot always distinguish between benign repository maintenance, release correction, policy-driven retagging, and malicious activity.
For this reason, our claims focus on the existence and nature of tag alterations and their reproducibility implications, rather than on attributing motive.

A related limitation is that the archival view yields lower bounds on temporal stability.
When we report how long a tag remained unchanged before a later alteration, we measure the minimum duration supported by the snapshots available to us, not necessarily the exact lifetime of the tag in the forge.
For the same reason, our counts of tag alterations should be interpreted as lower bounds.
This suffices to establish that substantial and potentially consequential alteration activity exists, but it may underestimate both true persistence and true alteration frequency.

\paragraph{Internal validity.}
A limitation of our archival-based approach is that less active repositories are archived less frequently by \SWH, leading to longer intervals between consecutive snapshots.
As shown in Table~\ref{tab:snapshot-frequency-stars}, observed snapshot frequency increases with repository popularity, meaning that we might capture less tag alterations from less popular repositories.
Even in high frequency snapshot repositories, we can miss alterations from tags that are pushed and subsequently overwritten between archival visits.
Likewise, because popular repositories are more likely to contain at least one tag, they are also more likely to exhibit at least one alteration, which may bias the proportion of impacted repositories shown in \Cref{fig:popularity_proportion}.
Such measurement bias could affect our correlation analysis with repository popularity: if low-activity repositories are undersampled relative to high-activity ones, we may underestimate the true alteration rate in less popular projects.
However, if low-popularity repositories are under-observed, the true ecosystem-wide prevalence of alterations is likely higher than what we report.

\paragraph{External validity.}
Our dataset is drawn from repositories archived in \SWH and from the subset of package definitions considered in the Nixpkgs cross-analysis.
This gives the study strong external validity because \SWH aggregates repositories across multiple forges, ecosystems, and project sizes, rather than reflecting a single platform's release practices.
At the same time, transferability is not uniform across all contexts: some environments may enforce stricter release processes, while others may rely more heavily on mutable symbolic references.
Likewise, Nixpkgs is especially informative for reproducibility research because it records expected hashes explicitly; ecosystems without comparable integrity metadata may face similar problems that are harder to detect.
Accordingly, we view Nixpkgs as a demonstrator of the phenomenon, not as a universal proxy for all package managers.

The GitHub Actions discussion should be interpreted in the same spirit.
We use it to illustrate a broader semantic tension around tags---between maintainers, who use mutable aliases, and consumers, who expect immutable dependency anchors---not to claim that all GitHub Actions usage is insecure or irreproducible by default.
The ecosystem is internally heterogeneous, and official guidance itself reflects that heterogeneity.
Our argument is that such ambiguity is precisely what makes tag mutability consequential for reproducibility and supply-chain integrity.

\paragraph{Conclusion validity.}
Finally, our study establishes that tag alterations occur at scale, that many moves change content, that the phenomenon intersects with real package metadata, and that our dataset captures an attack-shaped retargeting event consistent with the threat model we describe.
What we do \emph{not} claim is that every tag alteration is harmful, that every change to source content reflects malicious intent, or that tags are unsuitable for all release-engineering purposes.
Rather, our conclusion is narrower and evidence-based: tag names alone are insufficient as reproducible dependency anchors when downstream consumers require stable source identity over time.

 \section{Related work}
\label{sec:related}

Our work lies at the intersection of multiple bodies of previous work: software supply chain, reproducible builds, functional package management, software preservation, and version control system (VCS) alterations.
Prior work has studied these concerns separately, but the specific phenomenon of \emph{Git tag alteration} remains unexplored.

\paragraph{Software supply-chain and Reproducible Builds}

A broad body of work shows that modern software supply chains rely on deep dependency trees, externally hosted artifacts, and partially implicit trust relationships~\cite{ladisa2023supplychain, DBLP:journals/software/LambZ22, 10.1145/3641525.3663622}.
In that context, reproducible builds (R-B) have been introduced as a software integrity mechanism.
Lamb and Zacchiroli argue that they increase confidence that binaries correspond to auditable source code~\cite{DBLP:journals/software/LambZ22}.

Foundational and recent work on computational reproducibility emphasizes that results depend on the long-term recoverability of code, data, and procedures~\cite{881708, DBLP:journals/corr/abs-2011-10098}.
Related work on persistent scholarly objects similarly highlights the value of stable, citable software references~\cite{DBLP:journals/jodl/AustinBDKMNRSTV17}.
Our work complements that perspective by focusing on an earlier link in the chain: the identification of the source itself.
Even if downstream build steps are deterministic, reproducibility can still fail if the source is retrieved through a mutable symbolic reference such as a Git tag.
We show that Git tag names alone do not satisfy that requirement when tags are altered over time.

More broadly, recent systematizations of open-source software supply-chain attacks show that dependency resolution remains fragile when it relies on externally controlled infrastructure~\cite{ladisa2023supplychain}.
Git tag alteration fits this model: tags are socially treated as release identifiers, yet technically remain mutable.

\paragraph{Functional package management}

A second line of work studies how declarative specifications and functional package management~\cite{DBLP:conf/lisa/DolstraJV04} can make builds reproducible across machines and over time.
This literature provides an important baseline for our paper: reproducibility is achievable in practice when build inputs are controlled.

Recent empirical evidence shows that functional package management can enable reproducible builds at scale.
Malka et al.~demonstrate this on large package collections~\cite{DBLP:conf/msr/MalkaZZ25}, but that result should be interpreted as conditional on two alternative assumptions: continued availability of relevant artifacts in Nix binary caches \emph{or} continued stability of upstream source references.
Neither assumption is guaranteed over long time horizons.
If cached artifacts are eventually evicted (e.g., due to storage constraints), rebuilding may require re-fetching upstream sources; if those sources are referenced through mutable Git tags and have changed, Nix's hash check will detect the mismatch and the fetch/build will fail.
Thus, the issue is not silent substitution in Nix, but temporal fragility: a package reproducible today via cache can become unrebuildable later if the upstream tag no longer matches the recorded hash.

This perspective is reinforced by prior work on reproducibility ``through space and time''~\cite{DBLP:conf/icse/MalkaZZ24}, which ties reproducibility to reconstructing build environments long after the original build.
Our argument aligns with that temporal view, but shifts the focus to the stability of the \emph{source identifier}.
Relatedly, Vallet et al.~argue for practical, transparent, verifiable, and long-term reproducible research using Guix, emphasizing that reproducibility requires more than recording software names and version labels~\cite{valletPracticalTransparentVerifiable2022}.
A recent study similarly shows that long-term reproducibility depends on artifact longevity practices at the community level, including how artifacts are packaged and maintained over time~\cite{DBLP:conf/acmrep/GuilloteauCPGR24}.
Our findings support this claim: Git tags are convenient and human-readable, but they are not immutable content identifiers.

\paragraph{Software preservation}

Another closely related body of work examines how archival infrastructures support long-term reproducibility by preserving source code beyond the lifetime of its original hosting environment.
Prior work shows that source code archiving can rescue reproducible deployment when upstream-hosted artifacts disappear~\cite{10.1145/3641525.3663622}.

Our work addresses a different but complementary failure mode.
Here, the repository may remain available and the tag name may still exist, yet the content designated by that tag has changed.
In such cases, the problem is not just source disappearance but source ambiguity: a symbolic reference that once designated one commit later resolves to another.
This distinction matters because archival preservation alone does not eliminate the risk posed by mutable references.
If a workflow records only a tag name, even perfect archival coverage may still be insufficient to determine which source state was originally intended.

This also connects to broader concerns about scholarly reproducibility and the increasing dependence of scientific artifacts on code hosted on public forges.
Escamilla et al.~document the growing prevalence of Git hosting platform references in scholarly publications and the preservation challenges this creates~\cite{DBLP:conf/ercimdl/EscamillaKCRWN21}.
Follow-up work by the same line of research shows that many code references remain cited but not archived~\cite{DBLP:conf/icadl/EscamillaKCRWN23}.
While their work does not focus specifically on Git tags, it reinforces the importance of treating software references as long-lived research identifiers rather than transient hosting conveniences.

\paragraph{Version-control system alteration}

The Git literature written for practitioners also reflects this social expectation of stability.
The \emph{Pro Git} book states that ``A [...] tag is very much like a branch that doesn't change''~\cite{ProGit2014}.
Even though this wording is pedagogical rather than normative, it captures the widespread perception of tags as immutable release anchors.
Accordingly, the \texttt{git-tag} manual page warns that reusing a published tag name may leave different users with the same label on different commits, calling this practice ``insane'', and recommends issuing a new version identifier instead~\cite{git_tag_retagging_docs,torvalds_retagging_2007}.

The closest prior work to ours is the emerging literature on altered version control history in public repositories.
In particular, Rapaport et al.~\cite{DBLP:conf/kbse/RapaportPTZ25} show that public repository history can be modified after publication.
That work is an important precedent for ours, as it establishes that public version control state should not be assumed to be permanently stable.

Our work differs in two ways.
First, they examined altered histories broadly, whereas we focus specifically on \emph{tag alterations}---deletions and moves of references that downstream users and tools often treat as stable anchors.
Second, we focus on the consequences for \emph{software reproducibility and supply-chain integrity}.
Rewritten branches and commits matter for forensics and repository evolution; rewritten tags matter directly for dependency resolution, release identification, and rebuildability.

A related perspective comes from empirical studies of workflow reproducibility.
Grayson et al.~show that workflows may become irreproducible over time even when the workflow itself has not changed, due to changes in the surrounding software environment~\cite{DBLP:conf/acmrep/GraysonMKM23}.
This aligns with our longitudinal concern, but along a different axis: environmental drift rather than \emph{reference drift}.
Complementary studies focus instead on social and organizational mechanisms of reproducibility, including evidence from multiple years on artifact evaluation practices in systems and security venues~\cite{DBLP:conf/acmrep/OlszewskiLCBLBT25, DBLP:conf/acmrep/DEliaDGMPPPSSV25}, and research on whether citation dynamics reward reproducible work~\cite{DBLP:conf/acmrep/Raff23}.
Taken together with our findings, these results suggest that reproducibility over time depends on technical stability of source identification as well as community incentives and evaluation processes.

\paragraph{Positioning of our contribution}

In summary, prior work has shown that software supply chains are vulnerable to integrity failures, that functional package management can substantially improve build reproducibility, that long-term reproducibility depends on durable source and environment preservation, and that public version-control state can be altered after publication.
What remains underexplored is the role of Git tags as ambiguous objects that are simultaneously treated as release identifiers and mutable repository references.

Our contribution fills that gap.
We provide, to our knowledge, the first large-scale empirical study centered specifically on Git tag alterations, characterize the kinds of changes they induce, and connect them directly to reproducibility and software supply-chain risk through downstream package analysis and attack-oriented discussion.
In doing so, we argue that long-term reproducibility requires not only rebuildable environments and archived source availability, but also stable source \emph{identification}.

 \section{Conclusion}
\label{sec:conclusion}

Git tags are commonly viewed as convenient release identifiers and dependency anchors, yet Git does not guarantee that a tag name will continue to designate the same object over time.
In this paper, we conducted the first large-scale empirical study of tag alterations across public code, covering \OriginsWithMultipleSnapshotsShort Git repositories observed between \CutoffDate and \DatasetDate.

Across the repositories we analyzed, tag names are regularly moved or deleted, and many moves correspond to changes in the underlying source content rather than to metadata-only differences.
We further show that the phenomenon is not confined to obscure repositories: low-popularity projects dominate in absolute volume, while highly visible repositories exhibit the highest alteration prevalence and per-repository intensity.
Tag mutability is therefore a broad ecosystem phenomenon rather than an isolated anomaly.

In addition to measuring the phenomenon, our findings demonstrate direct consequences for reproducibility and software supply-chain integrity.
The cross-analysis with Nixpkgs shows that altered tags intersect with real package metadata, and that some packages still store hashes corresponding to earlier tag states that now resolve to different content.
Together with the attack case study we identified, these results show that mutable tags are not only a theoretical concern: they can silently invalidate assumptions about stable references and create opportunities for unintended or malicious retargeting.

The main implication of this work is that tag \emph{names} alone should not be treated as stable dependency anchors when reproducibility matters.
For build-critical inputs, immutable commit identifiers are the safer reference mechanism.
For computational research workflows in particular, dependency and artifact references should use immutable, content-addressed identifiers (such as SWHIDs) rather than mutable tags whenever possible, because they provide stronger integrity and long-term availability guarantees.
Platforms and tooling would benefit from better support for observing and auditing tag creation, movement, and deletion over time.
Recent GitHub introduction of immutable releases represents an initial step toward addressing tag mutability at the infrastructure level.
However, such features currently apply only to platform-specific release objects rather than to git tags in general, and require explicit opt-in by repository maintainers.

This work also opens several directions for future research.
First, the motivations behind tag alterations remain to be studied more directly, for instance through maintainer surveys or qualitative analysis of project practices.
Second, future work could investigate how different package ecosystems and build tools resolve and cache tags in practice, and how often this leads to silent drift.

We do not claim that every tag alteration is harmful, nor that tags are unsuitable for release engineering in general.
Rather, we show that mutable tag names and reproducible dependency resolution are in tension.
When downstream consumers require stable, time-invariant references, tags alone are insufficient.

 \begin{acks}
  The authors would like to thanks Jens Dietrich and Behnaz Hassanshahi for their
  invaluable suggestions that led us to start working on this project.
\end{acks}

\clearpage


\clearpage
\onecolumn
\appendix

\section{Appendix}
\label{sec:appendix}

\begin{table*}[!h]
\centering
\caption{Nix packages with build errors}
\label{tab:nixpkgs}
\resizebox{\textwidth}{!}{\begin{tabular}{|l|l|S|S|l|l|}
\hline
\textbf{Package}                      & \textbf{URL}                                   & \textbf{Tag}   & \textbf{GitHub stars} & \textbf{Category} & \textbf{Error}                                                                                                                  \\ \hline
\texttt{haunt}                        & \url{https://git.dthompson.us/haunt.git}       & \texttt{0.3.0} & Unknown               & \emph{deletion}   & \begin{tabular}[c]{@{}l@{}}error: cannot download source from any mirror\\ (when using \texttt{\small{--option substitute false --option substituters ""}})\end{tabular} \\ \hline
\texttt{python3Packages.e3-testsuite} & \url{https://github.com/AdaCore/e3-testsuite}  & \texttt{27.2}  & \num{6}                    & \emph{deletion}   & error: cannot download source from any mirror                                                                                   \\ \hline
\texttt{rita}                         & \url{https://github.com/activecm/rita}         & \texttt{4.8.1} & \num{1881}            & \emph{deletion}   & error: cannot download source from any mirror                                                                                   \\ \hline
\texttt{shelldap}                     & \url{https://github.com/mahlonsmith/shelldap}  & \texttt{1.5.1} & \num{23}                    & \emph{deletion}   & error: cannot download source from any mirror                                                                                   \\ \hline
\texttt{libv3270}                     & \url{https://github.com/PerryWerneck/libv3270} & \texttt{5.5.0} & \num{5}                     & \emph{move}       & error: hash mismatch                                                                                                            \\ \hline
\texttt{python3Packages.geometric}    & \url{https://github.com/leeping/geomeTRIC}     & \texttt{1.1}   & \num{80}                    & \emph{move}       & error: hash mismatch                                                                                                            \\ \hline
\texttt{superlu}                      & \url{https://github.com/xiaoyeli/superlu}      & \texttt{7.0.1} & \num{164}                   & \emph{move}       & \begin{tabular}[c]{@{}l@{}}error: hash mismatch\\ (when using \texttt{\small{--option substitute false --option substituters ""}})\end{tabular}                         \\ \hline
\end{tabular}}
\end{table*}

\begin{table*}[!h]
\centering
\caption{Archival snapshot frequency by GitHub star range}
\label{tab:snapshot-frequency-stars}
\begin{tabular}{|l|S|S|S|S|S|S|S|}
\hline
\textbf{Star range} & {\textbf{\#Unique Repositories}} & {\textbf{Mean}} & {\textbf{Q1}} & {\textbf{Median}} & {\textbf{Q3}} & {\textbf{P90}} & {\textbf{P95}} \\ \hline
\texttt{0--1}   & \num{103453} & \num{2.65}  & \num{0.78} & \num{1.45} & \num{2.76}  & \num{5.63}  & \num{8.60}  \\ \hline
\texttt{2--10}  & \num{19500}  & \num{4.79}  & \num{1.48} & \num{2.65} & \num{5.53}  & \num{10.14} & \num{13.59} \\ \hline
\texttt{11--100} & \num{16287} & \num{5.97}  & \num{2.04} & \num{3.95} & \num{8.31}  & \num{13.50} & \num{16.93} \\ \hline
\texttt{101--500} & \num{7037} & \num{9.56}  & \num{3.45} & \num{7.49} & \num{13.48} & \num{19.61} & \num{23.75} \\ \hline
\texttt{501+}   & \num{6348}   & \num{16.57} & \num{9.06} & \num{15.85} & \num{22.13} & \num{28.78} & \num{33.13} \\ \hline
\end{tabular}
\\[0.3em]
{All frequency values are in snapshots/year.}
\end{table*}

\end{document}